\documentclass[superscriptaddress,nofootinbib, amsmath,amssymb,preprintnumbers,prdfloatfix,twocolumn,PRL]{revtex4-2}
\usepackage{CJK}
\usepackage{graphicx}
\usepackage{dcolumn}
\usepackage{upgreek}
\usepackage{amsmath}
\usepackage{bm}
\usepackage[colorlinks=true,pdfstartview=FitV,breaklinks=true]{hyperref}
\usepackage[dvipsnames,table]{xcolor}
\hypersetup{urlcolor=BlueViolet,
	    citecolor=Plum,
	    linkcolor=PineGreen}
\usepackage{lipsum}	    
\usepackage{titlesec}
\usepackage{etoolbox}
\usepackage[normalem]{ulem}
\usepackage{tabularx}
\usepackage{amssymb}
\usepackage{float}	
\usepackage{caption}
\usepackage{subcaption}
\usepackage{comment}
\usepackage{appendix}
\usepackage{aasmacros}
\usepackage{subcaption}
\usepackage{afterpage}

 \usepackage{lettrine,Typocaps,Kinigcap,AnnSton,Kramer,Starburst}

\usepackage{relsize}

\begin{document}

\title{A dark matter hail:\\ Detecting macroscopic dark matter with asteroids, planetary rings, and craters}

 \author{Zachary S. C. Picker}
 \email{zpicker@physics.ucla.edu}

\affiliation{Department of Physics and Astronomy, University of California Los Angeles,\\ Los Angeles, California, 90095-1547, USA}

\begin{abstract}
\noindent Dark matter could be composed of macroscopic objects with large masses and geometric cross-sections spanning many decades. We investigate the potential interaction of such `stuff-sized' dark matter by considering its interactions with asteroids, planetary rings, and terrestrial bodies. This hail of dark matter could catastrophically destroy these Solar System objects, evaporate them from their orbits, or cause substantial cratering. We estimate these effects and use them to place competitive bounds on a wide, previously-unconstrained swathe of the dark matter parameter space.
\end{abstract}

\maketitle

\section{Introduction}

\lettrine[lines=5,findent=3pt]{M}{acroscopic}, composite, and ultraheavy objects are some of the oldest and most intriguing dark matter (DM) candidates. Early macroscopic dark matter candidates included primordial black holes~\cite{pbh,Hawking:1971ei,Carr:1974nx,Chapline:1975ojl,Carr:2020gox,Green:2020jor} (PBHs) and Standard Model quark nuggets~\cite{Witten:1984rs,Madsen:1986jg,Madsen:1998uh}, composite objects composed of up, down and strange quarks in a stable configuration and naturally formed in a strongly first order quantum chromodynamics phase transition. While one must turn to more worked scenarios to motivate quark nuggets today~\cite{Bai:2018dxf,Bai:2018vik,zhitnitsky_nonbaryonic_2003}, many more generic macroscopic dark matter scenarios have been studied on similar principles. In particular, a global U$(1)$ symmetry can stabilize large solitons consisting of fermions or scalars, known as Fermi balls~\cite{Lee:1986tr,Lee:1991ax,Kawana:2021tde,Hong:2020est,Hardy:2014mqa,Wise:2014jva,Wise:2014ola,Gresham:2017cvl,Chang:2018bgx,Flores:2020drq,Flores:2023zpf,DelGrosso:2023trq,Lu:2024xnb} and Q-balls~\cite{Coleman:1985ki,Kusenko:1997ad,Kusenko:1997si,Kusenko:2001vu,Kasuya:1999wu,Frieman:1988ut} respectively. 

Despite a continuing and even growing interest in macroscopic dark matter, there remains rather large gaps in the constraint space of dark matter with \textit{geometric} DM-nuclei cross sections and `object'-sized masses~\cite{cajohare}. Dark matter on the scale of bowling balls, fridges, whales, and pyramids have not been well-constrained. In particular, there are wide open gaps for dark matter at densities closer to that of nuclear matter, the expected density of quark nuggets and a useful benchmark for Q-balls and Fermi balls.

In this work we will investigate the consequences of generic macroscopic dark matter interacting with objects in our own Solar System. Since this dark matter scenario consists of relatively large chunks colliding at moderate frequencies with Solar System objects, we have dubbed this scenario a `dark matter hail' impacting planets, asteroids, and ring systems. Throughout we remain agnostic as to the specific dark matter model, merely specifying the dark matter radius $r$ and mass $m$. 

The advantage of looking in our Solar System is that there is a great wealth of high accuracy data on objects of widely varying scales, which have existed for very long time spans relatively undisturbed. Collisions of these dark matter chunks would leave craters on terrestrial planets, catastrophically destroy asteroids and particles in the rings of planets, or they could slowly transfer kinetic energy to larger objects until they reach escape velocity and are `evaporated' from their orbits. We estimate these processes here, providing strong, novel constraints on a wide swathe of previously-unconstrained parameter space for macroscopic dark matter in the `stuff-sized' region. 

\begin{figure*}[!ht]    \centering
    \includegraphics[width=.9\linewidth]{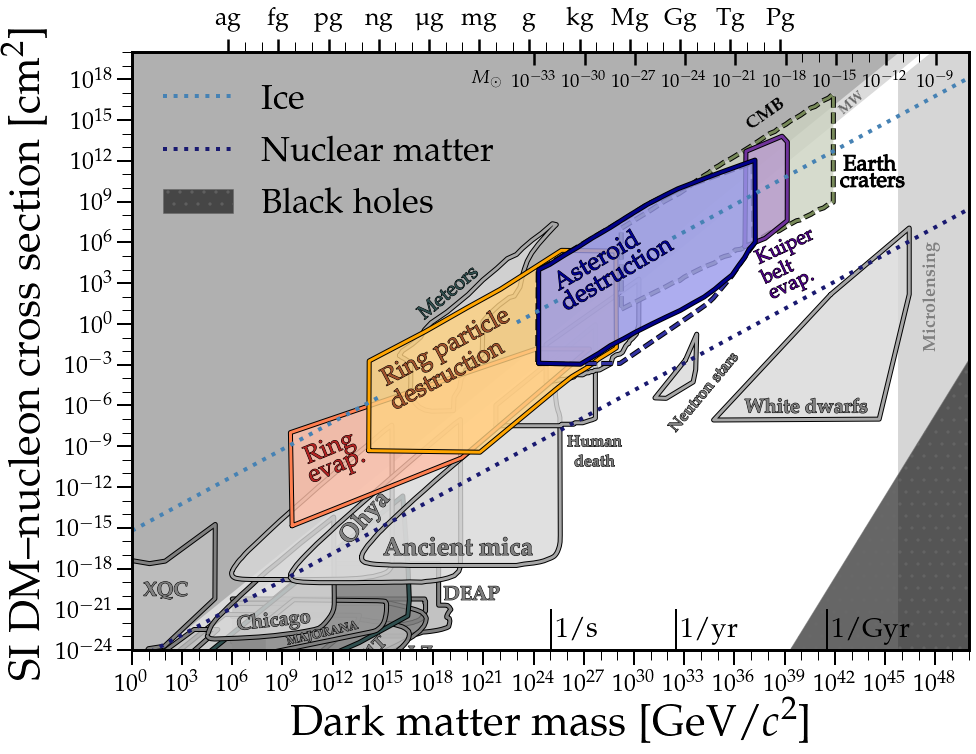}
    \caption{Constraints on the spin-independent dark matter--nucleon cross section as a function of dark matter mass. In {\color{Plum}indigo}, Kuiper belt objects above $\simeq40$km in radius are evaporated from their orbits around the sun by one or more collisions with the dark matter. In {\color{Blue}blue}, S-type asteroids are catastrophically destroyed by collisions with the dark matter. The dashed curve below shows the possible improvement in constraints if we observed small mass asteroids much older than their collisional lifetimes. In {\color{YellowOrange}yellow}, ice particles in Saturn's rings are destroyed by collisions, and in {\color{Bittersweet}salmon}, planetary ring particles are evaporated from their orbits around Saturn. The {\color{OliveGreen}olive--green} region with dashed border indicates where the dark matter creates more craters on Earth than expected from near-Earth objects, but cannot be confidently considered a constraint. For reference, the densities of ice, nuclear matter, and black holes are indicated on the plot. Existing constraints on dark matter are shown in {\color{Gray}grey}, as collected in Ref.~\cite{cajohare} and cited in Section~\ref{sec:disc}. On the bottom, ticks indicate how often dark matter of that size would impact the Earth. The largest previously-unconstrained region is on the larger mass end of this plot, spanning several orders of magnitude both horizontally and vertically. However, there is also a somewhat difficult-to-see region above the ancient mica constraints which were previously unconstrained~\cite{Cleaver:2025etu}.}
    \label{fig:constraints}
\end{figure*}

In section~\ref{sec:energy} we estimate the energy imparted onto an object by a macroscopic dark matter bullet which collides with it, either passing right through or becoming captured. In Section~\ref{sec:destruction} we estimate the destruction thresholds for dark matter collisions with asteroids and ring particles, and in Section~\ref{sec:evap} we estimate the evaporation of objects from their orbits by one or more dark matter collisions which do not destroy the object. In Sec.~\ref{sec:crater} we roughly estimate crater formation from dark matter impacts on the Earth or other terrestrial bodies, but ultimately argue that it cannot set competitive limits (without significantly more work). We conclude in Section~\ref{sec:disc}. The final constraints are shown in Fig.~\ref{fig:constraints} for those interested in the money plot.

\section{Energy transfer}\label{sec:energy}
\noindent First we need to estimate the effect of the collision between a dark matter bullet of mass $m$ and radius $r$ and a `regular' baryonic spherical object with mass $M$ and radius $R$. While in general smaller asteroids and ring particles are not necessarily spherical, we will make this approximation for the sake of tractability---more complex geometries will require detailed numerical simulation. We will also assume that the dark matter is itself not destroyed or otherwise majorly affected by the impact. 

When the dark matter collides with the object, it will elastically scatter with particles in a tube of radius $r$ as it passes through, until it is either stopped in the material or exits out the other side. As a simplified way to average over impact angles, we will take the tube length to be the average secant length in a sphere, $\frac43R$, so that the volume of this tube is $V=\frac43\pi R r^2$. This also sets a limit on the maximum dark matter radius $r$ such that the dark matter still passes cleanly through the object:
\begin{align}
r\leq \left(1-\sqrt{5}/3\right)R\simeq0.25R~.
\end{align}
Enforcing this ensures the geometry of the collision is sufficiently simple, and for the scenarios we consider here $r>0.25R$ occurs in regions of parameter space which are already ruled out by other dark matter constraints and can thus be safely ignored. In principal there is no reason that we cannot look at $r>0.25R$ to extend our limits upwards, but more work would be required for the unusual geometry there.

\begin{figure}[!ht]    \centering
    \includegraphics[width=.6\linewidth]{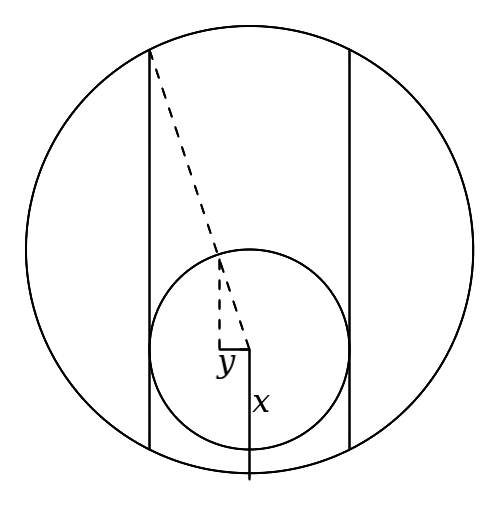}
    \caption{Diagram showing the dark matter passing through the object.}
    \label{fig:diagram}
\end{figure}

The number of collisions $N$ the dark matter experiences as it passes through the object is then given by $N\simeq \rho_o V/m_p$, where $\rho_o$ is the density of the object and where we use the proton mass $m_p$ as the approximate mass of a nucleon. We will assume the dark matter is colliding with a reasonably dense macroscopic object such that $N\gg1$. For simplicity let us assume that the particles in the object are stationary with respect to the dark matter and treat each collision as a linear elastic collision. The change in speed after one such collision is simply,
\begin{align}
    v_{DM,f}/v_{DM,i} = \frac{m-m_p}{m+m_p} ,
\end{align}
so that the change in overall velocity of the dark matter after $N$ collisions is,
\begin{align}\label{eq:X}
    v_{DM,N}/v_{DM} &= \left(\frac{m-m_p}{m+m_p}\right)^N \simeq e^{-X}~,\nonumber\\
    X&\equiv2Nm_p/m = 2\frac{r^2}{R^2}\frac{M}{m}~.
\end{align}
Then if $X>1$, the dark matter would have lost a significant portion of its velocity transiting the object, and if $X\gg1$, the dark matter can become captured within the object.
The energy $E$ imparted to the object by this collision can then be estimated by taking the difference between the kinetic energy of the dark matter before and after the impact. If the dark matter is captured, this is simply the total kinetic energy before the impact. If the dark matter is not captured, we then estimate,
\begin{align}\label{eq:energy}
    E \simeq 2 M\frac{r^2}{R^2} v_{DM}^2~.
\end{align}
Of course, some of this energy will be lost to ejected material directly displaced by the dark matter passing through the object. We will then estimate the efficiency of the energy transfer in the following way. Let us restore sphericity to the dark matter so that particles it collides with are moved in the direction of the normal to the sphere at that point, as illustrated in Fig.~\ref{fig:diagram}.

We will say that when the dark matter is at a distance $x$ throughout the tube, any particles found within the length $y$ from the center of the dark matter sphere will be ejected into space at the initial dark matter velocity.  This will only serve as an extremely rough estimate for the complicated physics of the real shock which would precede the dark matter. We assume that any material farther than $y$ will deposit their energy into the material of the object, contributing to its destruction. With some geometrical working, the volume of particles lost $V_{\rm lost}$ is given by,
\begin{align}
    V_{\rm lost} = \int_0^{4/3 R} \mathrm{d}x\pi y^2 = \pi r^3 \tan^{-1}(R/r)~,
\end{align}
so that the fraction of energy lost by this ejecta is simply,
\begin{align}
    E_\mathrm{lost}/E = \frac{9}{64}\frac{r}{R}\tan^{-1}\frac{R}{r}~.
\end{align}
When $r\ll R$, this is not a large fraction (as expected), but even as $r\rightarrow 0.25R$ this fraction only approaches $0.05$. For simplicity we will therefore ignore this minor correction in the following sections and assume that all the energy $E$ of the impact is transferred into the object via material shocks. If sufficient energy is transferred, the object can be destroyed, which we estimate in the following section.



\section{Catastrophic Destruction}\label{sec:destruction}
\noindent The study of the catastrophic destruction of asteroids has a long and rich history~\cite{holsapple_catastrophic_2019,benz_catastrophic_1999,ahrens_deflection_1992,melosh_asteroids_1997,melosh_dynamic_1992,housen_fragmentation_1990,bottke_asteroids_2002}. This is not only motivated by a desire to better understand how to avoid Earth-impacting events~\cite{noauthor_defending_2010,osinski_impact_2022,cooper_summary_1977}, but also to understand the history and formation of our Solar System~\cite{morbidelli_kuiper_nodate,MORBIDELLI2018262,bottkejr_fossilized_2005,bottkejr_linking_2005,davis_collisional_1997,pan_shaping_2005}. The generic criteria for asteroid destruction is quantified with the ratio $Q^*$, the kinetic energy of the impacter divided by the target mass. For catastrophic asteroid destruction, $Q^*_d$ is defined as the threshold above which two conditions hold: 1), the largest remaining fragment should be smaller than half the mass of the original; and 2), less than half of these fragments should eventually re-accumulate. We will adopt this criteria for destruction. 

There are two important regimes for asteroids. Below roughly $0.5-1$km, asteroids are considered to be in the material strength regime, where the primary forces which must be overcome to destroy them relate to the material strength of the rock or ice. Above this, the asteroids are gravity dominated, and the energy required to overcome the gravitational binding energy is significantly larger than the material strengths. 

However, destruction of gravity dominated asteroids is generally more difficult than merely overcoming the gravitational binding energy. Larger asteroids (from about $200$m to $10$km) are now thought to be predominantly rubble-pile asteroids~\cite{bagatin_how_2001}, gravitationally bound collections of rocks and boulders of varying sizes and compositions. Evidence for the ubiquity of rubble-pile asteroids are based on observations of low asteroid spins above about one kilometer in radius~\cite{warner_asteroid_2009,harris_shapes_2009}, direct measurements of densities and compositions of individual asteroids~\cite{veverka_nears_1997,fujiwara_rubble-pile_2006,pajola_evidence_2024}, the observed breakup of Shoemaker-Levy 9 over Jupiter~\cite{asphaug_density_1994,asphaug_size_1996,movshovitz_numerical_2012}, and the DART mission~\cite{daly_successful_2023,raducan_lessons_2024,raducan_reshaping_2022,jiao_sphdem_2023,jourdan_rubble_2023} which intentionally collided with the asteroid Dimorphos. Since rubble pile asteroids can be highly porous, the energy of the impacter is distributed inefficiently, resulting in larger $Q_d^*$ than a naive estimate which only requires the gravitational binding energy be overcome. In general, impacts on rubble-pile asteroids must be simulated numerically, and there is a high degree of variance in outcomes depending on the propagation of shocks through the specific geometry and composition of the rubble pile.

We will follow the review of ~\cite{holsapple_catastrophic_2019} who not only summarize the literature covering various analytical, empirical, and numerical calculations of this destruction, but provide two ultimate `best-fit' curves for $Q^*_d$. These are given both for rocky S-type asteroid targets and for the more porous C-type asteroid targets:
\begin{align}\label{eq:qd}
    Q_d^* = a\left(\frac{R}{R_a}\right)^\alpha + b\left(\frac{R}{R_b}\right)^\beta~,
\end{align}
with the parameters given in Table~\ref{tab:param}. The two summands account for the material strength regime and the gravity regime respectively.

\begin{table}[h]
    \begin{tabular}{| l | l | l | l | l | l | l |}\hline
    Asteroid type & $a$ [J/kg] & $b$ [J/kg] & $R_1$ [km] & $R_2$[km] & $\alpha$ & $\beta$ \\\hline
    S-type & $10^3$ & $10^6$ & $10^{-3}$ & $500$ & $-0.33$ & $1.65$ \\\hline
    C-type & $2\times10^3$ & $4\times10^5$ &  $10^{-3}$ & $500$ & $-0.25$ & $1.23$\\ \hline
    \end{tabular}
    \caption{$Q_d^*$ parameters for S-type and C-type asteroids from Ref.~\cite{holsapple_catastrophic_2019}.}
    \label{tab:param}
\end{table}

Although the density of asteroids is both size-dependent and has some variance amongst individuals, we used the benchmark values from Ref.~\cite{vernazza_vltsphere_2021}--- for S-type asteroids we took the density to be $3.0$ g cm$^{-3}$, and for C-types we used $1.7$ g cm$^{-3}$. For the particles in Saturn's rings we used the density of ice $0.92$ g cm$^{-3}$, but with the best-fit $Q_d^*$ for S-type targets---while these are of course not designed for a pure ice target, earlier studies~\cite{benz_catastrophic_1999} show that the destruction threshold for ice and basalt do not differ by more than a factor of approximately two, with ice being the easier of the two to destroy. 

The smallest observed asteroids are about $2$m large, such as TC25~\cite{reddy_physical_2016}---although this is not strictly an S- or C-type asteroid, we will use it as the benchmark smallest size. The largest S-type asteroids are Eunomia and Juno~\cite{vernazza_vltsphere_2021}, around $270$ and $250$km respectively, whereas the largest C-type are Hygiea at around $430$km and Ceres at $940$km~\cite{vernazza_vltsphere_2021}. However, our constraints are not very sensitive to the largest mass objects, since the destruction criteria ceases to be met before we reach such massive asteroids. 

The composition of Saturn's rings~\cite{miller_composition_2023,becker_characterizing_2016,french_saturns_2000} vary by subdivision, but generally the smallest observed particles are around a few mm and the largest on the order of $10$m. To the best of our knowledge, there is no current evidence for smaller particles than these in stable orbits~\cite{ohtsuki_size_2020,harbison_smallest_2013}, but if there was, the constraints here could be improved. We use $1$mm as the small-size cutoff for Saturn's ring particles, based on the treatment of monomers in Ref.~\cite{brilliantov_size_2015} which we will discuss in more detail in the following subsection.

Since $Q_d^*$ is generally smaller for S-type asteroids, they tend to provide marginally tighter constraints on the dark matter than the C-type asteroids. We thus ultimately use the S-type asteroid formula when deriving the constraints. Eq.~\ref{eq:qd} holds until the objects are smaller than about $1$ m, below which the curve takes a constant value. In Ref.~\cite{holsapple_catastrophic_2019} a cutoff on this estimate is imposed at $1$cm, but in order to consider the smallest particles in Saturn's rings, we will naively extend this to $1$mm. 

The threshold $Q_d^*$ is typically calculated for baseline values of $v=5.5$ km/s and with a $45^\circ$ impact angle. While the secant we are considering for our impact here is indeed angled with respect to the asteroid surface, we are considering dark matter velocities which are well in excess of this nominal value. To make the most precise constraints, we would need to numerically simulate the impact of the dark matter at these massive velocities with the asteroids, carefully calculating the material fragmentation and shock propagation. That is far beyond the scope of this work, so we will instead abuse the meaning of $Q_d^*$ and take it as a velocity-independent measure of the minimum energy needed to destroy the asteroid. However, as we will argue later, this may not be such a bad estimation, and in fact may even be conservative.

With this in hand, for an object of mass $M$, we only need to compare Eq.~\ref{eq:energy}, the energy $E$ imparted by the collision, with $MQ_d^*$ to find the minimum dark matter radius required to catastrophically destroy the object when $X<1$:
\begin{align}
    r\geq \frac{R}{\sqrt{2}}\left(\frac{v_{DM}}{\mathrm{m/s}}\right)(Q_d^*)^{1/2}.
\end{align}
When $X>1$, the condition for destruction becomes a lower bound for the dark matter mass $m$:
\begin{align}
    m \geq \frac{2}{v_{DM}^2} M Q_d^*.
\end{align}
Finally, in order to place constraints, we require that impacts happen more often than the known lifetime $\tau$ of a given object. The rate of collisions is simply,
\begin{align}
    t^{-1} \simeq n_{DM}\sigma v_{DM}~,
\end{align}
where we take the collision cross section to be $\sigma = \pi\left(R+r\right)^2\simeq \pi R^2$. For a given object of mass $M$, radius $R$, and lifetime $\tau$, if $t<\tau$ and the energy of the impact is sufficient to destroy it, we consider the dark matter constrained. These constraints are plotted in Fig~\ref{fig:constraints}.

\subsection{Ages of asteroids and ring particles}
\noindent The observed ages of asteroids and planetary rings must then be carefully considered in order to place constraints. For larger asteroids, observations of color~\cite{willman_asteroid_2011} consistently confirm ages in the billions of years, and indeed rubble-pile asteroids are very difficult to destroy by collisions with other asteroids~\cite{jourdan_rubble_2023}. For smaller asteroids, to the best of our knowledge, data on age is scarcer. We then must turn to simulations of the evolution of the asteroid belt, such as those in Ref.~\cite{bottkejr_linking_2005}. The smaller the asteroid, the easier they are to catastrophically destroy by collisions with other asteroids. By simulating the entire population, it is then possible to determine the average collisional lifetime of an asteroid of specific size, as given in Table~\ref{tab:lifetime}. We interpolate this data to determine the lifetime $\tau$ which should be compared against the destruction timescale $t$. 

\begin{table}[h]
    \begin{tabular}{| l | l | l | l | l | l | l |}\hline
    Diameter: [km] & $0.001$& $0.01$ & 0.1 & 1 & 10 & 100 \\\hline
    Collisional lifetime: [Myr] & 14 & 27 & 64 & 440 & 4700 & 34,000 \\\hline
    \end{tabular}
    \caption{Relation between radius of asteroid and collisional lifetime from Ref.~\cite{bottkejr_linking_2005}.}
    \label{tab:lifetime}
\end{table}

Of course, this lifetime is just the average lifetime of an asteroid of a given size. It is presumably the case that some of the small asteroids observed today are significantly older than the ages given here, which would substantially improve our derived dark matter constraints. To show this, the final constraint plot includes the region that would be constrained if we did indeed find small asteroids which could be proven to have survived $1$ Gyr without major disruptions. This is plotted as a dashed boundary in Fig.~\ref{fig:constraints}.

The age of Saturn's rings is another interesting question. We will assume that the rings are 100 million years old~\cite{dougherty_origin_2009,estrada_constraints_2023,iess_measurement_2019,teodoro_recent_2023}, although we note the caveat that this is generally a derived value based on a combination of observations and long timescale simulations of the ring evolution. However, the age of the rings does not necessarily imply that every ring particle of a given size is necessarily $100$ Myr old. In fact, it was shown in Ref.~\cite{brilliantov_size_2015} that the mass spectrum of the ring particles in Saturn's rings is well described by a population of `monomers' of a given size which collide with each other on day-to-week timescales. These collisions can be elastic, aggregating, or fragmenting, depending on the sizes and energies involved, causing larger ring particles to slowly be built up out of smaller ones, and more rarely destroyed again.  Amazingly, the equilibrium state of these processes for $1$ mm monomers accurately predicts the observed mass spectrum of the particles in Saturn's rings, including the upper cutoff at around $10$ m scales. 

In principle then, it is very difficult to determine a useful average collisional lifetime for a particle in Saturn's rings larger than the monomer size, since they are constantly growing and fragmenting by small amounts. However, the amount of time to grow such aggregates from scratch is presumably very long, since many monomers are required, and catastrophically destructive collisions of the largest particles will be proportionately rare. We thus choose to ignore this caveat and treat every particle in Saturn's rings as if they are indeed $100$ Myr old---this is certainly acceptable for the monomers themselves, which comprise the most interesting part of the constraint region here.

It is of course possible that even if $t<\tau$, the initial population of these objects might have been significantly larger and we are only today catching the `tail-end' of the destruction of these objects. However, such a scenario requires an exponentially fine-tuned initial population in order to match the correct abundances today, and so we will ignore this possibility.

We should also address one final caveat of this approach, and an opportunity for future work. The constraints here rely on the fact that if the asteroids or ring particles are catastrophically destroyed on sufficiently short timescales, that would trivially contradict existing observations of such objects. However, the criteria for destruction merely requires a significant amount of fragmentation, rather than complete obliteration. In reality, each collision would produce a population of smaller asteroids of varying sizes and velocities. In order to most accurately infer the effect of the dark matter on the population of asteroids or planetary ring particles, we would therefore need to run long timescale simulations of these populations. Such simulations already exist for the history of the asteroid belt~\cite{bottkejr_fossilized_2005,bottkejr_linking_2005}, the Kuiper belt~\cite{pan_shaping_2005,davis_collisional_1997}, and Saturn's rings~\cite{estrada_constraints_2023,durisen_large_2023}, and are important tools in understanding the history of the Solar System. The inclusion of a significant amount of dark matter collisions in these simulations could radically alter their conclusions, and by comparing these altered predictions to observation, we could obtain even tighter constraints on the dark matter, since the collective fragmentation processes from dark matter will produce overabundances of asteroids at small masses and underabundances at high masses which could be sensitively compared to observation.

\subsection{Conservativeness of estimate}
\noindent The catastrophic destruction estimate here may be conservative in two ways. Firstly, for the impacts between two asteroids (the primary historical use for $Q_d^*$) the bullet is itself a rocky object of similar density and composition to the target. This means that the a substantial amount of energy can be lost to the destruction and vaporization of the bullet itself. Secondly, the geometry of the shocks is quite different. Normally, the bullet deposits its energy on the edge of the asteroid in an area roughly equal to its radius. In our case, the bullet deposits its energy throughout the asteroid as it passes right through it.

For the first point, we can roughly quantify the energy lost to the vaporization of the bullet. For the sake of estimation let us assume that the standard bullet is composed of basalt. From~\cite{woskov_millimeter_nodate}, the total heat of vaporization of basalt is $H\simeq 25$kJ/cm$^{-3}$. It is then straightforward to calculate the ratio of the energy lost to vaporizing the bullet over the total kinetic energy of the bullet,
\begin{align}
E_\mathrm{vap.}/E_k = \frac{2H}{\rho_\mathrm{basalt}v^2}\simeq 0.6~,
\end{align}
where $\rho_\mathrm{basalt}=2.8$g cm$^{-3}$ and the standard bullet test speed is $v=5.5$km/s. The energy lost to vaporization (if completely vaporized) can therefore be a sizable fraction of the total kinetic energy of the impacter. In our scenario, in contrast, no energy would be lost to this vaporization.

Estimating the energy losses due to the geometry of the shocks is far more complicated, but we can get at least a rough sense of the estimation. One historically useful criteria~\cite{holsapple_catastrophic_2019,benz_catastrophic_1999} for asteroid destruction used the fact that the velocity of the material shocks dropped with the distance squared in the material as the shock spread from the impact point. If the bullet impacted on one side of the material, one could then calculate how fast the shock is moving on the opposite side of the asteroid, at a distance $2R$, and if the velocity at that distant point was still larger than the escape velocity, then the asteroid could be considered disrupted. In our case, the energy is not just deposited at the impact site, but throughout the secant the dark matter bullet carves out. The most distant point from the average secant we considered here is perpendicular from its midpoint, at a distance approximately $1.75R$. Then the velocity of the shock at this point will be $1.3$ times the velocity of the shock at $2R$. If we instead considered a trajectory through the middle of the asteroid (rather than the average secant), the velocity would be four times larger. Combined with the previous estimate of the energy lost to vaporization, then, we could be underestimating the constraints by an order of magnitude or so.

\section{Orbit evaporation}\label{sec:evap}
\noindent We also considered the possibility that the dark matter might not catastrophically destroy the object, but could instead transfer sufficient momentum to it so that it becomes unbound in its orbit. We will refer to this as the `evaporation' of the object from its present orbit. 

If objects in our Solar System were moving through an isotropic background of dark matter, then there would be a consistent bias towards head-on collisions, regardless of the direction of motion of the object with respect to the Solar frame of reference. This would cause the objects to on average experience a drag force, provided the momentum transfer is sufficiently small in each collision. However, this situation does not occur for the objects in consideration here---the Solar System is moving through the galaxy at a rate approximately equal to the dark matter velocity, causing the dark matter background to be extremely anisotropic. 

While this dark matter `wind' could be ignored in the previous consideration of destruction (where only one collision is necessary to place constraints), it can no longer be ignored here. Not only is the most probable collision always occurring in the direction of the wind (rather than the direction of motion of the object itself), but the relative velocity in the headwind is significantly higher and so momentum transfer is larger. In the frame of reference of the galaxy, this will look like a drag force on objects in our Solar System, whereas in the Solar frame, it will look like the objects are being accelerated in the direction of the wind. It is important to remember that the orbital plane of the planets in our Solar System is not aligned with the axis of the Milky Way, being tilted around $60^\circ$ with respect to the Milky way plane. The dark matter wind is currently coming from the direction of the constellation Cygnus, at a declination close to $40^\circ$.

The Solar System orbits the galaxy with a period of approximately 230 million years. Thus on timescales of $\mathcal{O}(10~\rm Myr)$ or less we can treat the dark matter wind as a force acting on the object in a constant direction. Over longer timescales, the direction of the dark matter wind rotates through $2\pi$ as the Solar System orbits the Milky Way. If the dark matter collisions constituted a constant, continuous force on any object, this force would average out to zero over the course of one galactic orbit. As we will see, however, when collisions with the dark matter are infrequent enough to be discrete events (as is the case for macroscopic dark matter), the expected change in velocity after one galactic orbit is larger than zero. The intuition for this observation is to imagine that the collisions are changing the object's velocity in a `random walk.' In a random walk, the average final position remains at the origin, (over many trials), but the\textit{ average displacement} goes like the square root of the number of steps.

It is beyond the scope of this paper to treat the dark matter wind in full details, including the dark matter velocity distribution, careful integration over long timescales, and the resulting distribution of final velocities of the objects. We intend to treat this more thoroughly in forthcoming works. To make a simple order-of-magnitude estimate here, we will merely calculate the expected change in velocity after one full galactic orbit. We will then take that acceleration and use it as the average acceleration, even over non-integer numbers of orbits around the Milky Way. This will give us a rather conservative estimate of the final velocities of the objects, since over partial orbits of the galaxy, the directions of dark matter collisions are more aligned and so would in fact accelerate the object at a higher rate. 

Ultimately we want to know how many collisions $k$ are required to increase the velocity by a factor $\delta$. For evaporation, escape velocity at a given radius is only a factor $\delta=\sqrt{2}$ larger than the speed of circular orbit, so $\delta=\sqrt{2}$ will be the key threshold. We will use this threshold as a conservative way to place constraints, since the evaporation of objects from their orbits would trivially contradict observations. However, it is certainly the case that we could place much more sensitive constraints than this---even a small disruption to the trajectories of Saturn's rings, for instance, should have large observable consequences. We will leave this more sensitive computation to future work, noting that even our more crude constraints eliminate much of the meaningful parameter space available.

The number of collisions over some timespan $\tau$ is simply,
\begin{align}\label{eq:k}
    k = \frac{\rho_{DM}}{m}\sigma v_{\rm rel} \tau~,
\end{align}
where we again take the cross section $\sigma = \pi\left(R+r\right)^2\simeq \pi R^2$ and where $v_{\rm rel}$ is the relative velocity of the dark matter wind, which will assume for simplicity to be $v_{\rm rel} = 2v_{\rm DM}$.

First let's examine the case when $X<1$ and the dark matter bullets are passing through the objects. By conservation of momentum we can write the final velocity after these collisions as,
\begin{align}
    \vec{v}_f = \vec{v}_o + v_{rel}\frac{m}{M}\left(1-e^{-X}\right)\sum_k \hat{r}_i~,
\end{align}
Where $\vec{v}_o$ is the object's initial velocity. To account for the change in the dark matter wind, $\hat{r}_i$ are random unit vectors taken from the set of directions the dark matter wind points over one full galactic orbit (i.e., a two-dimensional circle). We are simplifying here by assuming that all collisions are happening in the direction of the dark matter wind, since it is by far the most probable case. A full calculation which averages over all collisions in three dimensions is beyond the scope of this simple estimation.

We will then make use of a handy result for the expected value of the magnitude of the sum of $k$ random, isotropic unit vectors~\cite{103170}. When $k$ is large, for vectors in $\mathbb{R}^d$ the expected value of this magnitude approaches,
\begin{align}
    \langle \Big|\Big| \sum_k \hat{r}_i \Big|\Big|\rangle \simeq \sqrt{\frac{2k}{d}}\frac{\Gamma\left(\frac{d+1}{2}\right)}{\Gamma\left(\frac{d}{2}\right)}~,
\end{align}
which evaluates to $\simeq 0.89\sqrt{k}$ in two dimensions, as we require here. 

Before proceeding, it is important to point out the effect of this square root on the calculation. One might naively expect that the evaporation of Solar System bodies should be independent of dark matter mass, since a smaller mass implies less momentum transfer but proportionately more collisions. However, averaging over the isotropy of the dark matter and the resulting random walk of the object's momentum is responsible for this $\sqrt{k}$ growth which ultimately breaks the degeneracy in dark matter mass. As the dark matter mass decreases, we easily see that $\vec{v}_f-\vec{v_o} \propto \sqrt{m}\rightarrow0$ and we recover the required result that a continuous force rotated around $2\pi$ will not have impacted the final velocity of the object. Of course, it would be interesting to consider what the actual displacement of the object would look like over one full (or partial) orbit of the galaxy---this could lead to more sensitive constraints which we leave to a more detailed forthcoming analysis of this effect.


Using the reverse triangle inequality we can then solve for a lower bound on $k$:
\begin{align}
    \sqrt{k} &\geq \frac{1}{0.89}\frac{M}{m}\frac{v_o}{v_{\rm rel}}\frac{\delta-1}{\left(1-e^{-X}\right)}~.
\end{align}
\begin{align}
    &\simeq \frac{1}{0.89}\frac{v_o}{v_{\rm rel}}\frac{R^2}{r^2}\frac{\delta-1}{2}~.
\end{align}
Where the second line takes $X<1$. Substituting Eq.~\ref{eq:k} allows us to estimate a lower limit on the dark matter radius such that the orbit will be eventually evaporated:
\begin{align}\label{eq:rlowevap}
    r^2 \geq \frac{R}{0.89}\left(\frac{m}{\pi \rho_{DM}v_{\rm rel}\tau}\right)^{1/2}\frac{v_o}{v_{\rm rel}}\frac{\delta-1}{2}~.
\end{align}

Now let us return to the regime when $X>1$. For simplicity let us assume that in this limit the dark matter is captured in the object and all of the momentum is transferred. Let us also assume that this a perfectly inelastic collision, with no losses from cratering or ejecta. The calculation follows similarly to the previous section, with the final velocity now being given by,
\begin{align}
    \vec{v}_f = \frac{1}{M+km}\left(M\vec{v}_o + mv_{\rm rel}\sum_k \hat{r}_i~\right),  
\end{align}
and in this limit the dark matter mass must be added to the total object mass after each collision. We again use the result for the expected value of the sum of unit vectors and apply the triangle inequality, finding,
\begin{align}
    \frac{\sqrt{k}\frac{m}{M}}{\delta\left(1+(.89^2)k\frac{m}{M}\right)-1}\geq \frac{1}{0.89}\frac{v_o}{v_{\rm rel}}~,
\end{align}
which does not have a beautiful solution for $k$. We can proceed by noting that for the objects we have in question, $(.89^2)km<M$ is very easy to satisfy, since this limit can be simplified to,
\begin{align}
    R>2.9\times10^{-6}~\mathrm{cm}\left(\frac{\rho_o}{1\mathrm{gcm}^{-3}}\right)^{-1}\left(\frac{\tau}{1\mathrm{Gyr}}\right)~,
\end{align}
and the smallest objects we will consider here are $\simeq1$mm ice particles in planetary rings. Then we have,
\begin{align}
    \sqrt{k} \geq \frac{1}{0.89}\frac{v_o}{v_{\rm rel}} \frac{M}{m}(\delta-1)~.
\end{align}
This allows us to place a lower bound on the dark matter mass which would evaporate the given objects,
\begin{align}\label{eq:mlowevap}
    m\geq \frac{16\pi}{9(0.89^2)}\left(\delta-1\right)^2\frac{v_o^2}{v_{\rm rel}^3}\frac{R^4}{\tau}\frac{\rho_o^2}{\rho_{DM}}~.
\end{align}
If we have $k>1$ and Eqs.~\ref{eq:rlowevap}~and~\ref{eq:mlowevap} are satisfied for $\delta=\sqrt{2}$, we can consider the dark matter scenario constrained. For consistency, we should also insist that the object is not trivially destroyed in this process, using the calculations of the previous section. We will assume that if one collision does not destroy the object, then many subsequent ones will not either---this may not be a safe assumption, but if the object was to be destroyed by multiple collisions, then we could anyways still consider the scenario constrained.

In principal, there exists constraints for all the previously considered populations of objects. However, the constraints for evaporation in the asteroid belt do not improve on the catastrophic destruction constraints for two reasons. Firstly, small asteroids have rather short collisional lifetimes before they are destroyed. Secondly, the orbital velocity of asteroids is in the range of $20$ km/s, which is somewhat higher than more distant objects from the sun---the strongest evaporation constraints come from light particles with already low orbital velocities.

As it turns out, the most interesting large objects to consider are Kuiper belt objects. We did not discuss Kuiper belt objects in the previous section since they offered no real advantage over the S-type asteroid destruction constraints. Here however the key advantage is the comparatively low orbital speed of around $5$ km/s. This allows even larger Kuiper belt objects to be evaporated within the lifetime of the Solar System. Kuiper belt objects have been observed with radii spanning approximately $1-1000$ km, the largest of which is Pluto~\cite{pan_shaping_2005}. Importantly, Kuiper belt objects above $40$ km or so in radius are thought to be pristine remnants of the formation of the Solar System. For the lifetimes of Kuiper belt objects, we follow Ref.~\cite{pan_shaping_2005}, who infer the collisional lifetimes of differently-sized Kuiper belt objects from the location of the kink in their mass spectrum.

These constraints could be improved again by considering Oort cloud objects~\cite{jewitt_icy_2009,rickman_oort_2014}, which would have have yet smaller orbital speeds than Kuiper belt objects. However, since observations of the Oort cloud are even today limited to indirect observations of comet trajectories, these limits will not be included here. 

Saturn's rings again turn out to be interesting, since they have a range of orbital velocities spanning roughly $2-25$ km/s~\cite{miller_composition_2023}. The strongest constraints thus come from the lower part of this range, in the outer rings. One issue here is that as previously discussed, the particles in Saturn's rings are constantly colliding with each other and thus distributing the extra energy they would get from the dark matter collisions. However, since all the particles should be `heating up' roughly at the same rate, the effect of many self-collisions should ultimately not stop the total mass of the rings from evaporating. Again, a more precise constraint would require long timespan numerical simulations of the evolution of Saturn's rings, including the incident dark matter population.

The constraints for both the Kuiper belt objects and the evaporation of Saturn's rings are plotted in Fig.~\ref{fig:constraints}. For clarity, the constraints for the evaporation of the asteroid belt, other ring systems, or other large bodies are not shown, since they are superseded by the plotted constraints.

\section{Earth cratering}\label{sec:crater}
\noindent Now we investigate constraints on macroscopic dark matter from the frequency of collisions with Earth and the formation of craters. The rate of impacts on Earth for a given dark matter mass have also been indicated on Fig.~\ref{fig:constraints} with ticks along the bottom axis. While one might be concerned that these rates seem quite high, it ultimately is not easy to constrain the dark matter using Earth impacts. These constraints should be expected to overlap significantly with the asteroid constraints, since one of the first methods historically for determining if an asteroid was destroyed was to consider the size of craters---if the crater would be larger than the asteroid, it was considered destroyed~\cite{benz_catastrophic_1999,holsapple_catastrophic_2019}. 

To begin, we can use the approximate empirical results of Ref.~\cite{hughes_approximate_2003} as an estimate for the radius of a crater on Earth formed from a given impact energy:
\begin{align}\label{eq:crater}
    E = 1.6\times10^{17}~\mathrm{J}~\left(\frac{\ell}{\mathrm{km}}\right)^{2.59}~,
\end{align}
where $\ell$ is the radius of the crater. Of course there is a wide variance in cratering when considering factors like the ground material, density of the meteor, and the angle and velocity of impact, so this estimate should be taken only as a useful guide for order-of-magnitude estimations. 

There is a similar complication here to the asteroid destruction estimation, since small and massive dark matter impacters will pass very deep into the earth, rather than depositing all their energy at the impact site. The volume of the material that the dark matter bullet passes through is now just $V=\pi r^2h$, where $h$ is the depth the dark matter passes into the Earth. Then as in Eq.~\ref{eq:X}, we can define $X\equiv 2\pi r^2hm$, so that if $X>1$ at a particular depth $h$, we can consider all of the energy having been imparted into the target material. 

For simplicity let us consider the crater shape to be a half-sphere. Then if $X>1$ when the depth is smaller than the crater radius, i.e. $h<\ell$, all of the kinetic energy of the dark matter is deposited into the Earth and is used up in the formation of the crater. It is then trivial to use Eq.~\ref{eq:crater} to determine the dark matter mass required to create a crater of a given size $\ell$.

When $X<1$, however, only some of the energy is deposited into the surface material of the Earth. The rest of the energy of the dark matter is deposited deeper underground. The seismic effects of this impact is well beyond the scope of this paper, but we can still approximate the crater size. There will be a specific depth at which the energy imparted into the ground up to that point would create a crater of exactly that width. This is straightforward to solve for, ultimately leading to a lower bound on the dark matter radius which creates a crater of a given size:
\begin{align}
    r \geq 0.89~\mathrm{km}~\left(\frac{\ell}{km}\right)^{0.80}\left(\frac{\rho_{\oplus}}{3\mathrm{gcm}^{-3}}\right)^{-1/2}\left(\frac{v_{DM}}{220\mathrm{km/s}}\right)^{-1}~,
\end{align}
where $\rho_{\oplus}$ is the density of the Earth's surface material. To estimate the constraints, however, we need \textit{empirical} data about the rate of asteroid collisions with Earth. In general, this is extremely complicated, because the active geological history of the Earth makes detailed evidence scant. While we can learn a significant amount from the hundreds-or-so of known craters~\cite{osinski_impact_2022}, estimating the overall rate of impacts from these surviving craters is more difficult, since one would need to model the number of events which \textit{didn't} lead to craters discoverable today. Such modeling is beyond the scope of this work, and possibly an impossible challenge given the rich geological history of Earth. This makes it very difficult to place confident constraints on macroscopic dark matter purely from the numbers of craters on active geological bodies like Earth.

\begin{figure}[!ht]    \centering
    \includegraphics[width=\linewidth]{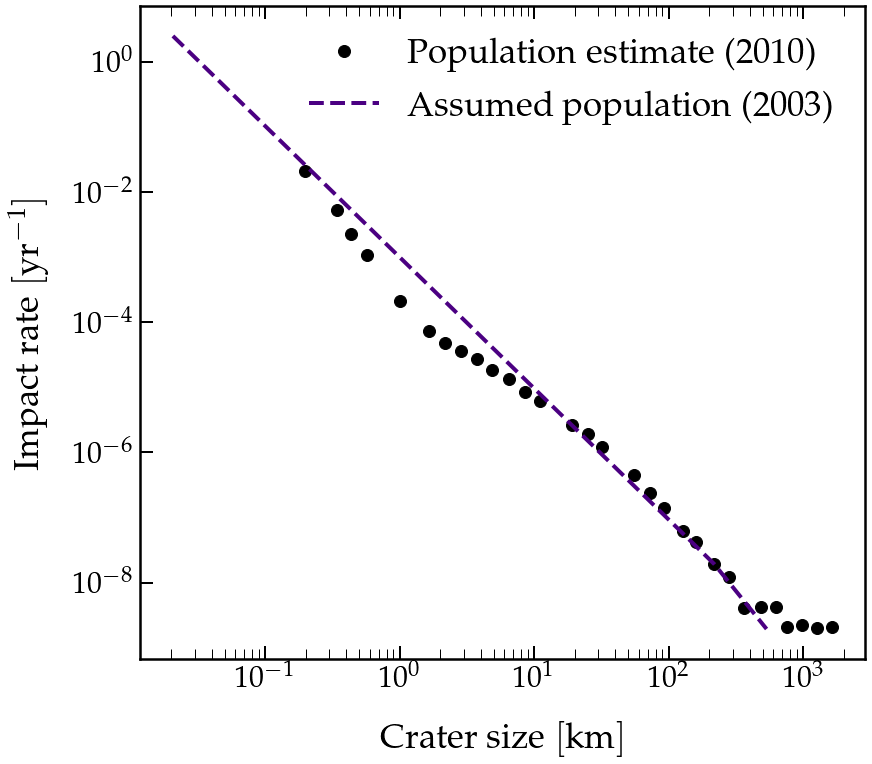}
    \caption{Relation between crater size and impact rate on Earth, from data in Refs.~\cite{noauthor_defending_2010}~and~\cite{hughes_approximate_2003}. The filled dots represent an estimate of the population which would lead to these impacts, with an approximate curve for the assumed population shown in the dashed {\color{Plum}indigo} line. }
    \label{fig:crater}
\end{figure}

Still, in order to get a sense of where these constraints could lie, we can compare the macroscopic dark matter crating rate to the \textit{expected} cratering rate from known near-Earth bodies. Data on the frequency and impact energy of meteors on Earth was taken from Fig.~2.4 in Ref.~\cite{noauthor_defending_2010} and combined with the approximate empirical formula relating crater size on Earth, Eq.~\ref{eq:crater}, in order to estimate the expected timescales of cratering on Earth, which we plot in Fig.~\ref{fig:crater}.

On Fig.~\ref{fig:constraints} we show the region where collisions producing a particular crater size from macroscopic dark matter would be more frequent than the equivalent estimates from near-Earth objects. We also sensibly impose $r\leq \ell$. For impact rates which are small compared with geological timescales (approximately thousands to millions of years), one could presumably place reliable constraints on the dark matter, since the consequences of such an increased cratering rate would likely be observationally evident---this region anyway overlaps with the asteroid destruction constraints. For large timescales, however, no constraint can be confidently made, since we strictly require empirical evidence which would definitively show that these collision rates are inconsistent with Earth's geologic history. Because of this uncertainty, we show the constraints with dashed outline on Fig.~\ref{fig:constraints}.

Constraints from craters could be improved in a few ways beyond the scope of the present work. Firstly, it would be more helpful to use a geologically inactive body, such as the moon, where craters can last for billions of years. The rate of moon cratering is indeed widely studied~\cite{liakos_neliota_2024,yang_lunar_2020,MORBIDELLI2018262,Gallant:2006us}, but we face similar problems to the Earth, in that the region which is best constrained will overlap considerably with the asteroid destruction region. It is unlikely that surveys of moon craters will help us push into the higher mass regime, since it will again be difficult to distinguish very massive dark matter craters from regular cratering---particularly when accounting for heavy uncertainties around the late bombardment epoch. The best hope for improvement in the constraints would be to carefully study the morphological difference between dark matter craters and regular meteor craters, as was done in the case of primordial black hole cratering on terrestrial bodies~\cite{Yalinewich:2021fdr,Santarelli:2025eye}. 

One interesting way to improve the constraints could be to carefully account for the dark matter wind. As discussed earlier, the Solar System is at the moment moving in the direction of the Cygnus constellation, at around $40^\circ$ declination. This means that the North pole of the Earth, as well as many other bodies whose spin axes are approximately perpendicular to the plane of the Solar System, would see a larger flux of dark matter impacts at present day than their South poles. If there is no difference in the present cratering rate across the North and South poles of objects like the moon, stronger constraints on the dark matter could perhaps be placed. This is somewhat stymied by the fact that the dark matter wind will be averaged out over $230$ Myr timescales, so these constraints would need to focus on timescales closer to the order of $10$ Myr.

\section{Results and conclusions}\label{sec:disc}

\noindent If the dark matter is composed of macroscopic objects, it could manifest as a `hail' impacting on objects in our Solar System. This work presents order-of-magnitude estimates for using a number of Solar System bodies---asteroids, planetary rings, and terrestrial cratering---in constraining macroscopic dark matter. Our new dark matter constraints are plotted along with existing constraints in Fig.~\ref{fig:constraints}. Previous constraints come from a wide variety of sources, spanning cosmology, astrophysics, geology, and more traditional dark matter direct detection. The constraints included on this plot (some of them hidden behind new constraints) include the cosmic microwave background (CMB)~\cite{Dvorkin:2013cea,Gluscevic:2017ywp}, Milky Way satellites (MW)~\cite{Nadler:2019zrb}, microlensing~\cite{Niikura:2017zjd,SinghSidhu:2019tbr}, white dwarves~\cite{Graham:2018efk}, neutron stars~\cite{SinghSidhu:2019tbr}, human death~\cite{SinghSidhu:2019loh}, fireballs~\cite{SinghSidhu:2019cpq}, meteor radar data~\cite{Dhakal:2022rwn}, ancient mica~\cite{Acevedo:2021tbl}, Ohya~\cite{Bhoonah:2020fys}, Skylab~\cite{Bhoonah:2020fys}, Chicago~\cite{Cappiello:2020lbk}, DEAP~\cite{DEAPCollaboration:2021raj}, XQC~\cite{Bhoonah:2020dzs},  DAMA~\cite{Bernabei:1999ui,Bhoonah:2020dzs}, CRESST~\cite{Bhoonah:2020dzs}, CDMS~\cite{Bhoonah:2020dzs}, XENON1T~\cite{XENON:2023iku}, and LZ~\cite{LZ:2024psa}. These constraints, along with the plotting code, were compiled and published by Ciaran O'Hare in Ref.~\cite{cajohare}. 

Our bounds here ultimately cover many decades of previously-unconstrained parameter space, demonstrating significant promise in continuing to use the Solar System as a testing ground for dark matter. Ultimately, we would like to place constraints (or open avenues for discovery) down to the densities of nuclear matter, where we expect quark nuggets, Fermi balls and Q-balls to preferentially reside.  Besides being cool to study, the rich history of the Solar System and its fascinating planets, moons, and asteroids may be a fertile testing ground for some of these most intriguing dark matter candidates.

There are a number of interesting ways to improve these constraints which could be the subject of future work. Firstly, it would be ideal to include the effect of dark matter destruction in long time-span numerical simulations of systems like the asteroid belt, the Kuiper belt, or Saturn's rings. This could allow a more careful tracking of fragmented pieces and orbital changes and thus more sensitive comparisons to observations of these systems today. Secondly, numerical simulations of asteroid destruction and terrestrial cratering by the dark matter collisions will be necessary to place more stringent constraints. Beyond this, constraints can be improved with further astronomical observations---in particular, finding older and smaller asteroids will increase the asteroid destruction bounds. Finally, we could look to detailed analyses of the cratering history of the moon and other geologically inactive objects like Ganymede or Mercury~\cite{Santarelli:2025eye} to search for the possible effect of a distinct population of macroscopic dark matter craters. Accounting properly for the dark matter wind may be a significant help, since if the cratering rate is sufficiently rapid, there should be a discrepancy in the number of recent craters of a particular size in the direction of the dark matter wind. It may also be necessary to distinguish the morphologies of dark matter craters from regular impact craters.

\section*{Acknowledgements}
\noindent 
This work was partially written on Gadigal, Chumash and Tongva land. I would like to thank Aaron Vincent, Ciaran O'Hare, and Alexander Kusenko for useful discussions on this somewhat-goofy topic. This work was supported by the U.S. Department of Energy (DOE) Grant No. DE-SC0009937. 
This work made use of N\textsc{um}P\textsc{y}~\cite{numpy2020Natur.585..357H}, S\textsc{ci}P\textsc{y}~\cite{scipy2020NatMe..17..261V}, and M\textsc{atplotlib}~\cite{mpl4160265}.

\bibliographystyle{bibi}

\bibliography{main.bib}

\end{document}